\documentclass[preprint]{jpsj2}

\title{Nonequilibrium Steady States and Fano-Kondo Resonances 
\\ in an AB Ring with a Quantum Dot}

\author{\textsc{Junko Takahashi}\thanks{E-mail address: 
junchi@ruri.waseda.jp}, \textsc{Shuichi Tasaki}}
 
\inst{Department of Applied Physics, School of Science and Engineering, 
\\ Waseda University, Tokyo 169-8555}

\abst{Electron transport through a strongly correlated quantum dot (QD) 
embedded in 
an Aharonov-Bohm (AB) ring is
investigated with the aid of the finite-$U$ slave-boson mean-field (SBMF) 
approach 
extended to nonequilibrium regime.
A nonequilibrium steady state (NESS) of the mean-field Hamiltonian is 
constructed with the aid of the C$^*$-algebraic approach for studying 
infinitely extended systems. 
In the linear response regime,
the Fano-Kondo resonances and AB oscillations of the conductance 
obtained from the SBMF approach
are in good agreement with those from the numerical renormalization 
group technique (NRG) by Hofstetter {\it et al. } by using twice larger Coulomb interaction.
At zero temperature and finite bias voltage, the resonance peaks
of the differential conductance tend to split into two.
At low bias voltage, the split of the asymmetric resonance can be 
observed as an increase of the conductance plateau. 
We also found that the differential conductance has zero-bias 
maximum or minimum depending on the background transmission via
direct tunneling between the electrodes.
}

\kword{Fano-Kondo resonance, nonequilibrium steady states, 
AB oscillation, quantum dot, AB ring, finite-$U$ slave-boson 
mean-field approach, C$^*$-algebra}

\begin{document}
\maketitle
\newpage

\section{Introduction} 
Coherent transports of electrons through a quantum dot (QD) embedded in 
an Aharonov-Bohm (AB) ring have been widely studied in recent years. 
For such systems,
interference between the localized states in the 
QD and the continuous states in the direct scattering channel
causes asymmetric resonance peaks in the conductance \cite{Kobayashi,Kobayashi1},
the so-called Fano resonances \cite{Fano}. 
On the other hand,
strong Coulomb interaction induces the Kondo effect, i.e.,
the formation of a singlet bound state among localized 
and conduction electrons. 
In case of transport through single QD, it is seen as a 
plateau in the resonance tunneling peak, 
where the conductance reaches the unitary limit value ($2e^2/h$). 
Thus, when the Coulomb interaction is strong enough, 
an interplay between the Fano and Kondo effects is expected in
the AB ring with a QD.

In the linear response regime, the interplay
was theoretically studied by 
Hofstetter {\it et al.},
employing the numerical renormalization group (NRG) technique \cite{NRG},
and they predicted 
suppression of the Kondo plateau by the Fano 
resonance (the Fano-Kondo resonance). Such a behavior
has been observed by recent experiments\cite{Katsumoto,Aikawa}.
They also studied the AB oscillations in the plateau regime \cite{NRG}
and showed that the phase of the AB-oscillation is almost fixed at 
$\pi /2$ \cite{ABphase} 
and its frequency is almost doubled, both relative to
the AB-oscillation at higher gate voltage.

Outside the linear response regime, the conductance 
maximum at zero bias voltage \cite{Kondoeffect1,Kondoeffect2}
characterizes the Kondo effect in the transport through a single QD.
The conductance drop at finite bias voltage is considered to be 
due to inelastic scattering involving electrons with energy 
between the chemical potentials of the two electrodes in case of infinite Coulomb repulsion.
An AB ring with a QD is predicted to have a conductance minimum or maximum
at zero bias voltage depending on the magnetic field \cite{Bulka}. 
In spite of these researches, we think that the nonequilibrium properties of an
AB ring with a QD particularly in case of large but finite Coulomb interaction
are not fully understood partly because of the lack
of complete characterization of nonequilibrium ensembles.

On the other hand, the method of C$^*$-algebra has been widely applied to 
rigorous investigations of nonequilibrium steady states (NESSs) of infinitely
extended quantum systems\cite{Ojima1,Ojima2,Ruelle1,Ruelle2,HoAraki,Jaksic1,Jaksic2,Jaksic3,ST1,ST2,Frolich}.
The method was originally developed to deal with equilibrium statistical mechanics\cite{Bratteli,RuelleEq,RuelleEq1}
and, then, found to be useful also for studying nonequilibrium 
properties\cite{Ojima1,Ojima2,Ruelle1,Ruelle2,HoAraki,Jaksic1,Jaksic2,Jaksic3,ST1,ST2,Frolich}. 
Particularly, Fr\"ohlich, Merkli and Ueltschi\cite{Frolich} have shown the existence 
of NESS  
and its characterization for junction systems, which consist of several 
free-fermionic 
reservoirs mutually interacting via bounded interactions.
Their results cover the AB ring with a QD in case that the electrodes 
are free fermion systems,
but explicit characterizations of NESS are not given.

In this paper, we investigate the transports in an AB ring with a 
single-level QD, employing the finite-$U$ slave-boson mean-field (SBMF) 
approach of Kotliar and Ruckenstein \cite{Kotliar} extended 
to nonequilibrium steady states.
The NESS with respect to the mean-field Hamiltonian is constructed
based on the C$^*$-algebraic approach as in the noninteracting 
case\cite{JPSJ1,JPSJ2}.
The infinite-$U$ 
\cite{Ramon,Aono1,Aono2,Aono3,WU,Dong5,Kang,Wu,Aharony} and 
finite-$U$ \cite{Dong1,Dong2,Dong3,Dong4,Dong6, Dong7,Jing,Guo,Gustavo} 
SBMF approaches in combination with the Keldysh Green function method
are successfully applied to the study of strongly correlated systems.
In particular, the main features of the Kondo effect in the transport 
through a single QD  
such as the conductance plateau and the zero-bias maximum 
have been well described by the finite-$U$ SBMF 
approach \cite{Dong2,Dong7}.
In contrast to these works, we focus on the finite-$U$ case and 
treat the mean-field NESS in terms of the C$^*$-algebraic approach,
which leads to a global characterization of NESSs and, we believe, has 
an advantage compared to the Keldysh method in this respect.

The rest of this paper is arranged as follows.
In Sec.\ref{Sec2}, after describing the model of the AB ring with a QD
and reviewing the slave-boson description of Kotliar and Ruckenstein,
the nonequilibrium SBMF approximation is explained.
Relations among mean-fields and Lagrange multipliers are derived 
from the equations of motion
of the auxiliary bosons within Dirac's formulation of constrained 
dynamics in contrast to the original SBMF approach \cite{Kotliar}, which is
based on the saddle-point approximation of the path integral 
representation of the free energy.
The self-consistent equations are closed by evaluating the averages
of certain fermionic variables with respect to a NESS for the mean-field
Hamiltonian, which is constructed with the aid of the C$^*$-algebraic
approach. 
Section \ref{Sec3} is devoted to the summary of the self-consistent equations
and the derivation of the average current. Within this approximation, 
the Coulomb interaction is shown to result in the renormalization of 
the dot energy and the reduction of the dot level-width.
In Sec.\ref{Sec4}, we study the conductance in the linear response regime
and show that (i) \ the Fano-Kondo resonances 
and AB oscillations
obtained from the SBMF approach are in good agreement 
with the results from the NRG technique by employing twice large Coulomb energy\cite{NRG}.
In addition, the SBMF approach is shown to give a simple view 
to the underlying physical processes.
In Sec.\ref{Sec5}, we study the differential conductance at fixed
average number of dot electrons.
We have found that (ii) \ the Fano-Kondo resonances are deformed by 
the bias voltage so that the resonance peaks have the tendency to 
split into two peaks and that (iii) \ the differential conductance
may have zero-bias maximum or minimum depending on the transmission
via the direct tunneling between the electrodes.
The former is due to the narrowing of the dot level-width by the 
Kondo effect and, at low bias voltage, can be observed as
an increase of the conductance plateau, which is consistent with
the recent observation by Katsumoto {\it et. al. } for the T-shaped QD systems \cite{Katsumoto}. 
Section \ref{Sec6} is devoted to concluding remarks. 
In Appendix, we explain the details of the construction of asymptotic field operators necessary 
for the construction of NESS.  
 
\section{Nonequilibrium Slave-Boson Mean-Field Approach}\label{Sec2}
\subsection{Model and slave-boson approach}
Our system is described by the Anderson single impurity 
Hamiltonian~\cite{Anderson} with 
the direct channel 
between left and right electrodes:
\begin{equation}
H=H_L+H_R+H_D+H_T+H_{LR}. \label{total}
\end{equation} 
The first and second term are Hamiltonians of the two-dimensional 
electrodes
\begin{eqnarray}
H_{r}=\sum _{\sigma}\int 
dk\omega _{kr}a_{k\sigma r}^{\dagger }a_{k\sigma r }
\qquad
(r =L \ {\rm or } \ R)
\end{eqnarray}
where $\omega _{kr }$ is the single-electron energy: 
$\omega _{kL}=\omega _{kR}-eV=\hbar^2 k^2/(2m)-eV/2$ with $V$ the bias voltage 
and
$m$ the effective mass, 
and $a_{k\sigma r}^{\dagger } (a_{k\sigma r})$ is the creation 
(annihilation) operator
of the electrode electron with wave number $k$ and spin $\sigma $. 
The third term $H_{D}$ describes the single-level dot with the Coulomb 
interaction $U$:
\begin{eqnarray}
H_D=\sum _{\sigma}\epsilon _{\sigma }c_{\sigma }^{\dagger }c_{\sigma }
+Un_{\uparrow }n_{\downarrow }
\end{eqnarray}
where
$c_{\sigma }^{\dagger } (c_{\sigma })$ is the creation (annihilation) 
operator of electrons in the dot, 
$n_{\sigma}=c_{\sigma}^{\dagger }c_{\sigma}$ the number operator 
and $\epsilon _{\sigma}$ 
the single-level energy. 
The tunneling 
between the dot and electrodes is given by
\begin{equation}
 H_{T}=\sum _{\sigma}\int dk [u_{kL} a_{k\sigma L}^{\dagger }c_{\sigma }+u_{kR} a_{k\sigma R}^{\dagger }c_{\sigma }+({\rm H.c.})] \label{original H}
\end{equation}
and the transmission through the direct channel by
\begin{equation}
H_{LR}=\sum _{\sigma}\iint dkdq [Wu_{kL}u_{qR}e^{i\varphi }
a_{k\sigma L}^{\dagger }a_{q\sigma R}+({\rm H.c.})] \ , 
\end{equation}
where the tunneling matrix elements $u_{kL}$, $u_{kR}$ and 
$Wu_{kL}u_{qR}$ (hence $W$) are taken to be real.
Because the resonance shapes do not strongly depend on the functional 
form of the tunneling 
matrix elements, one can assume that the interaction Hamiltonian $H_{LR}$ 
is separable
as above without loss of generality.  
The AB phase $\varphi $ is related to the magnetic flux $\Phi$ penetrating the ring
via $\varphi = e\Phi /(\hbar c)$ with $c$ the velocity of light.  

In order to examine the transport through the strongly correlated QD,
we employ the 
finite-$U$ slave-boson mean-field (SBMF) approach proposed by
Kotliar and Ruckenstein~\cite{Kotliar}. 
In order to describe the doubly occupied state as a bosonic single 
particle excitation, they
have introduced four auxiliary bosonic fields ${\hat e}, {\hat p}_{\sigma }
\ (\sigma=\uparrow, \downarrow)$ and $\hat d$, 
respectively, labeling
the empty, singly occupied 
and doubly occupied states,
and a fermionic field $f_{\sigma}$, describing the fermionic feature
of the QD electrons.
The expression of the Hamiltonian (\ref{total}) in terms of
the new variables is obtained by replacing the original QD electron
operators as follows:
$$
\begin{matrix}
c_\sigma^\dag c_\sigma &\to &f_\sigma^\dag f_\sigma \cr
n_\uparrow n_\downarrow &\to &{\hat d}^\dag {\hat d} \cr
c_\sigma &\to &{\hat z}_\sigma f_\sigma \nonumber
\end{matrix}
$$
where ${\hat z}_\sigma$ is a bosonic variable defined by
\begin{equation}
{\hat z}_{\sigma}=(1-{\hat d}^{\dagger}{\hat d}-{\hat p}_{\sigma}^{\dagger}
{\hat p}_{\sigma})^{-1/2}({\hat e}^{\dagger}
{\hat p}_{\sigma}+{\hat p}_{\bar \sigma}^{\dagger}{\hat d})
(1-{\hat e}^{\dagger}{\hat e}-{\hat p}_{\bar \sigma}^{\dagger}
{\hat p}_{\bar \sigma})^{-1/2}.  
\label{KRst1}
\end{equation}
The additional degrees of freedom simplify the Coulomb interaction
$Un_\uparrow n_\downarrow$ as $U{\hat d}^\dag {\hat d}$, but, at the same time, 
introduce vectors not describing physical states. 
Then, physical vectors $\lvert \Phi_{ph} \rangle$ are picked up by 
a set of constraints 
\begin{align}
{\hat \phi}_{1\sigma}|\Phi_{ph}\rangle \equiv&({\hat p}_{\sigma}^{\dagger}
{\hat p}_{\sigma}+{\hat d}^{\dagger}{\hat d}-f_{\sigma}^{\dagger}f_{\sigma})\lvert 
\Phi_{ph} \rangle=0
\notag \\ 
{\hat \phi}_2|\Phi_{ph}\rangle \equiv&({\hat e}^{\dagger}{\hat e}+
{\hat d}^{\dagger}{\hat d}+
\sum _{\sigma}{\hat p}_{\sigma}^{\dagger}{\hat p}_{\sigma}-1)\lvert \Phi_{ph} \rangle=0  
\label{constraint}
\end{align}
As easily seen, the subspace of the enlarged Hilbert space defined by 
(\ref{constraint}) is equivalent to the original Hilbert space.
Note that, since ${\hat \phi}_{1\sigma }, {\hat \phi}_{2}$ and $H$
commute with each other, no further constraints are necessary.

As discussed in detail by Dirac~\cite{Dirac,Dirac1,Dirac2}, 
corresponding to the constraints ${\hat \phi}_{1\sigma}|\Phi_{ph}
\rangle=0$ and ${\hat \phi}_{2}|\Phi_{ph}
\rangle=0$,
$q$-number Lagrange multipliers
${\hat \lambda}_{\sigma }^{(1)}$ and ${\hat \lambda}^{(2)}$ 
should be introduced to the Heisenberg equation of motion
and, with respect to the effective Hamiltonian
${{\widetilde H}\equiv~H+\sum_\sigma~{\hat \lambda}_{\sigma }^{(1)}
{\hat \phi}_{1\sigma}
+{\hat \lambda}^{(2)}{\hat \phi}_{2}}$,
any dynamical variable $A$ satisfies the standard equation of motion 
as a state equation
\begin{eqnarray}
{dA\over dt}|\Phi_{ph}\rangle={i\over \hbar}[{\widetilde H},A]
|\Phi_{ph}\rangle
\label{EQofMOTIONst}
\end{eqnarray}
where $|\Phi_{ph}\rangle$ is an arbitrary physical state
and the explicit expression of ${\widetilde H}$ is 
\begin{align}
{\widetilde H}=&H_L+H_R+H_{LR}
+\sum_{\sigma}
\int dk [{\hat z}_{\sigma}^{\dagger }(u_{kL}
f_{\sigma}^{\dagger}a_{k\sigma L}+
u_{kR}f_{\sigma}^{\dagger}a_{k\sigma R})+(H. C. )]
+U{\hat d}^{\dagger}{\hat d}
\notag \\
&+\sum_{\sigma}\epsilon_{\sigma}f_{\sigma}^{\dagger}f_{\sigma}
+\sum_{\sigma}\lambda_{\sigma}^{(1)}(
{\hat p}_{\sigma}^{\dagger}{\hat p}_{\sigma}+
{\hat d}^{\dagger}{\hat d}-f_{\sigma}^{\dagger}f_{\sigma})
+\lambda^{(2)}\bigl(
{\hat e}^{\dagger}{\hat e}+{\hat d}^{\dagger}{\hat d}+
\sum_{\sigma}{\hat p}_{\sigma}^{\dagger}{\hat p}_{\sigma}-1\bigr)
\ .
\end{align}
Note that, under the constraints (\ref{constraint}), the equation of
motion (\ref{EQofMOTIONst}) is equivalent to the Heisenberg equation
of motion with respect to the original Hamiltonian (\ref{total}).

\subsection{Nonequilibrium mean field approximation}

The key idea of the SBMF approximation is to replace all the slave-boson
operators and the Lagrange multipliers by their average values, which 
will be denoted as $e, p_\sigma, d, \lambda_\sigma^{(1)}$
and $\lambda^{(2)}$, and to impose, instead of (\ref{constraint}),
weaker constraints among averages:
\begin{align}
&\lvert e \rvert^{2}+\lvert d \rvert^{2}+
\sum _{\sigma} \lvert p_{\sigma} 
\rvert^{2}=1 
\label{WeakerConstraint1} \\
&\langle f_{\sigma}^{\dagger}f_{\sigma}\rangle_{ss}=\lvert d \rvert^{2}
+\lvert p_{\sigma} \rvert^{2}  \ ,
\label{WeakerConstraint2}
\end{align}
where $\langle f_{\sigma}^{\dagger}f_{\sigma}\rangle_{ss}$ stands for an 
average of $f_{\sigma}^{\dagger}f_{\sigma}$.
This approximation can be extended to nonequilibrium 
steady states (NESSs) by regarding $e, p_\sigma, d, \lambda_\sigma^{(1)},
\lambda^{(2)}$ and $\langle f_{\sigma}^{\dagger}f_{\sigma}\rangle_{ss}$
as the NESS averages with respect to the mean field effective 
Hamiltonian:
\begin{equation}
H_{eff}= H_{F}+
U|d|^2+\sum _{\sigma}\lambda_{\sigma}^{(1)}
(|p_{\sigma}|^2+|d|^2)
+\lambda^{(2)}\bigl( |e|^2+|d|^2+
\sum_{\sigma}|p_{\sigma}|^2-1\bigr)
\end{equation}
where $H_{F}$ is the fermionic part:
\begin{equation}
H_{F}=
H_L+H_R+H_{LR}
+\sum_{\sigma}\int dk [z_{\sigma}^*(u_{kL}f_{\sigma}^{\dagger}a_{k\sigma L}+
u_{kR}f_{\sigma}^{\dagger}a_{k\sigma R})+(H. C. )]
+\sum_{\sigma}{\bar \epsilon}_{\sigma}f_{\sigma}^{\dagger}f_{\sigma} \ ,
\end{equation}
${\bar \epsilon}_\sigma=\epsilon_\sigma-\lambda_{\sigma}^{(1)}$
is the effective energy of the QD level and the mean value $z_\sigma$ is 
obtained from (\ref{KRst1}) by replacing the bosonic operators to their
averages.
The characterization and properties of NESS with respect to $H_{eff}$
will be discussed in the next subsection.

We have seven averages to be determined and the constraints provide three
conditions. Hence, four more conditions are necessary.
Originally, Kotliar and Ruckenstein~\cite{Kotliar} derived them by the 
saddle-point approximation for the equilibrium free energy, but this 
approach can not be applied to NESS.
Here, we derive those conditions from the equations of motion
of the slave boson fields.
For example, $\hat d$ obeys the equation of motion:
\begin{align}
i\hbar {d\over dt} {\hat d}=-\sum _{\sigma } \int dk 
\{ [{\hat z}_{\sigma }^{\dagger },{\hat d}](u_{kL}f_{\sigma }^{\dagger }
a_{k\sigma L}
+u_{kR}f_{\sigma }^{\dagger }&a_{k\sigma R}) 
+[{\hat z}_{\sigma},{\hat d}](u_{kL}a_{k\sigma L}^{\dagger }
f_{\sigma }+u_{kR}a_{k\sigma R}^{\dagger }f_{\sigma }) \} \notag \\
&+(U+\sum _{\sigma }\lambda _{\sigma}^{(1)}+\lambda ^{(2)}){\hat d} \ . \label{eq:d}
\end{align}
\noindent
By replacing the slave-boson operators and Lagrange multipliers to their 
mean values and by evaluating NESS averages of the fermionic operators,
one gets
\begin{eqnarray}
\sum _{\sigma}\left( \frac{\partial z_{\sigma}^{*}}{\partial d^{*}}M_{\sigma }
+\frac{\partial z_{\sigma }}{\partial d^{*}}M_{\sigma}^{*}\right)+(U+\sum_{\sigma }
\lambda_{\sigma }^{(1)}+\lambda^{(2)} )d=i\hbar{d\over dt}d=0 
\label{SelfST1}
\end{eqnarray}
where $M_{\sigma}=\int dk \langle u_{kL}f_{\sigma}^{\dagger}a_{k\sigma L}+
u_{kR}f_{\sigma }^{\dagger}a_{k\sigma R}\rangle_{ss}$.
Here, the average of 
$[{\hat z}_\sigma,{\hat d}]$ is evaluated as
\begin{align}
[{\hat z}_\sigma,{\hat d}]&= \
[(1-{\hat d}^{\dagger}{\hat d}-{\hat p}_{\sigma}^{\dagger}
{\hat p}_{\sigma})^{-1/2},{\hat d}] \
({\hat e}^{\dagger}
{\hat p}_{\sigma}+{\hat p}_{\bar \sigma}^{\dagger}{\hat d})
(1-{\hat e}^{\dagger}{\hat e}-{\hat p}_{\bar \sigma}^{\dagger}
{\hat p}_{\bar \sigma})^{-1/2} \cr
&= \sum_{n=1}^\infty{(2n-1)!!\over (2n)!!}
\ [({\hat d}^{\dagger}{\hat d}+{\hat p}_{\sigma}^{\dagger}
{\hat p}_{\sigma})^n,{\hat d}] \
({\hat e}^{\dagger}
{\hat p}_{\sigma}+{\hat p}_{\bar \sigma}^{\dagger}{\hat d})
(1-{\hat e}^{\dagger}{\hat e}-{\hat p}_{\bar \sigma}^{\dagger}
{\hat p}_{\bar \sigma})^{-1/2}
\cr
&= -\sum_{n=1}^\infty{(2n-1)!!\over (2n)!!}\sum_{l=0}^{n-1}
({\hat d}^{\dagger}{\hat d}+{\hat p}_{\sigma}^{\dagger}
{\hat p}_{\sigma})^{n-1-l}{\hat d}
({\hat d}^{\dagger}{\hat d}+{\hat p}_{\sigma}^{\dagger}
{\hat p}_{\sigma})^l
({\hat e}^{\dagger}
{\hat p}_{\sigma}+{\hat p}_{\bar \sigma}^{\dagger}{\hat d})
(1-{\hat e}^{\dagger}{\hat e}-{\hat p}_{\bar \sigma}^{\dagger}
{\hat p}_{\bar \sigma})^{-1/2}\cr
&\to
-\sum_{n=1}^\infty{(2n-1)!!\over (2n)!!}n 
(|d|^2+|p_{\sigma}|^2)^{n-1}d
(e^* p_{\sigma}+p_{\bar \sigma}^*d)
(1-|e|^2-|p_{\bar \sigma}|^2)^{-1/2}
=-{\partial z_\sigma \over \partial d^*}
\nonumber
\end{align}
and that of $[{\hat z}_\sigma^\dag,{\hat d}]$ as 
$\displaystyle -{\partial z_\sigma^* \over \partial d^*}$.
Similarly, from the equations of motion of $\hat e$ and ${\hat p}_\sigma$,
one has
\begin{align}
&\sum _{\sigma}\left( \frac{\partial z_{\sigma}^{*}}{\partial e^{*}}M_{\sigma }
+\frac{\partial z_{\sigma }}{\partial e^{*}}M_{\sigma }^{*}\right)
+\lambda ^{(2)}e=0 
\label{SelfST2} \\ 
&\sum _{\tau}\left( \frac{\partial z_{\tau}^{*}}{\partial p^{*}_{\sigma}}
M_{\tau}
+\frac{\partial z_{\tau }}{\partial p^{*}_{\sigma }}M_{\tau }^{*}\right)
+(\lambda _{\sigma }^{(1)}+ \lambda ^{(2)} )p_{\sigma}=0 
\label{SelfST3}
\end{align}

When one neglects the Zeeman effect as in this article,
a straightforward investigation shows that $e, p_\sigma, d$ and $z_\sigma$
can be taken as real positive and, then, $M_\sigma$ is real.
Therefore, by setting $d=\sqrt D,\;  p_{\sigma }=\sqrt P_{\sigma },\;  
e=\sqrt E$, the constraints (\ref{WeakerConstraint1}), (\ref{WeakerConstraint2})
and the conditions (\ref{SelfST1}), (\ref{SelfST2}), (\ref{SelfST3})
cast into self-consistent equations for $D$ and $\lambda^{(1)}_\sigma$:
\begin{align}
&z_\sigma={\sqrt{EP_\sigma}+\sqrt{DP_{\bar \sigma}}\over
\sqrt{(1-D-P_\sigma)(1-E-P_{\bar \sigma})} } 
\label{selfconsistent1} \\
&\lambda ^{(1)}_{\sigma}=2\sum_{\tau }
\left(\frac{\partial z_{\tau }}{\partial E}
-\frac{\partial z_{\tau}}{\partial P_{\sigma}}\right)M_{\tau}
\label{selfconsistent2} \\
&2 \sum_{\sigma }\frac{dz_{\sigma}}{dD}M_{\sigma }+U=0 
\label{selfconsistent}
\end{align}
where $\displaystyle {d\over dD}={\partial \over \partial D}
+{\partial \over \partial E}-\sum_\tau
{\partial \over \partial P_\tau}$
and one should set
$P_\sigma=\langle f_\sigma^\dag f_\sigma\rangle_{ss} -D$, 
$E=D+1-\sum_\sigma \langle f_\sigma^\dag f_\sigma\rangle_{ss}$
after evaluating all the derivatives.

\subsection{Nonequilibrium steady states for the mean field Hamiltonian} 

Now we turn to the averages of fermionic operators which are evaluated
at a nonequilibrium steady state (NESS) with respect to the fermionic part
$H_F$ of the effective Hamiltonian. Since we are interested in transports
through mesoscopic contacts, we consider a NESS such that, very far
from the contact, each electrode recovers equilibrium with individual 
temperatures and chemical potentials. According to the C$^*$-algebraic
approaches\cite{Ruelle1,Ruelle2,Jaksic1,Jaksic2,Jaksic3,Frolich}, an average of an observable
$A$ with respect to such a NESS is can be evaluated as
\begin{equation}
\langle A \rangle_{ss}=\lim_{t\to +\infty} \langle e^{i{H_F\over \hbar}t}A
e^{-i{H_F\over \hbar}t} \rangle_0
\end{equation}
where $\langle A \rangle_0\equiv {\rm Tr}(\rho_{l}A)$ is the average 
with respect to a local equilibrium state 
$\rho_{l}$\cite{fnote1}:
\begin{equation}
\rho_{l}={1\over \Xi} e^{-\beta_L(H_L-\mu_L N_L)} 
e^{-\beta_R(H_R-\mu_R N_R)} \ .
\end{equation}
In the above, $\Xi$ is the normalization constant, $\beta_r$,
$\mu_r$ and $N_r$ ($r=L,R$) are, respectively, the
inverse temperature, chemical potential and the total number of 
particles:\break 
$N_r\equiv \sum_\sigma \int dk \ a_{k\sigma r}^\dag 
a_{k \sigma r}$ of each electrode.

The existence and characterization of the NESS averages can be heuristically
explained in terms of the M\o ller operator:
$\widehat \Omega ={\displaystyle \lim_{t\to -\infty}} e^{iH_Ft/\hbar}
e^{-iH_0t/\hbar}$ where $H_0=H_L+H_R$. Indeed, since the initial density
matrix $\rho_l$ commutes with $H_0$, one has
$$
\langle A \rangle_{ss}=\lim_{t\to +\infty} \langle e^{-iH_0t/\hbar} 
e^{i{H_Ft/\hbar}}A
e^{-i{H_Ft/\hbar}}e^{iH_0t/\hbar} \rangle_0
=\lim_{t\to -\infty} \langle e^{iH_0t/\hbar}
e^{-i{H_Ft/\hbar}}A
e^{i{H_Ft/\hbar}}e^{-iH_0t/\hbar} \rangle_0
\ .
$$
This seems to suggest $\langle A \rangle_{ss}=\langle \widehat \Omega^\dag A \widehat \Omega 
\rangle_0$, but, as in the derivation of the Gell-Mann Low 
relation\cite{Fetter}, 
the change of normalization should be taken into account 
and one obtains
\begin{equation}
\langle A \rangle_{ss}={\langle \widehat \Omega^\dag A \widehat \Omega 
\rangle_0 \over
\langle \widehat \Omega^\dag \widehat \Omega \rangle_0} \ .
\end{equation}
According to the rigorous approaches\cite{Ruelle1,Ruelle2,Jaksic1,Jaksic2,Jaksic3,Frolich}, 
the existence of $\langle A \rangle_{ss}$ for an aribtrary finite
observable $A$ requires the absence of bound states of $H_F$, which
is assumed throughout this article.
In this case, $\widehat \Omega \widehat \Omega^\dag ={\bf 1}$ and $\widehat \Pi\equiv 
\widehat \Omega^\dag \widehat \Omega$ is a projection onto the ortho-complement of the 
states involving dot electrons, namely,
one should have $\widehat \Pi=(1-f_\uparrow^\dag f_\uparrow)
(1-f_\downarrow^\dag f_\downarrow)$\cite{fnote2}.

Simple characterization of the NESS is given in terms of the asymptotic
fields. According to the scattering theory\cite{scattering}, 
the M\o ller operator $\widehat \Omega$ generates an `in' field:\break
$\beta_{k\sigma \lambda}^\dag \equiv
{\displaystyle \lim_{t\to -\infty}}e^{iH_Ft/\hbar}e^{-iH_0t/\hbar}
a_{k\sigma \lambda}^\dag e^{iH_0t/\hbar}e^{-iH_Ft/\hbar}=
\widehat \Omega a_{k\sigma \lambda}^\dag \widehat \Omega^\dag$, which
creates `in' states from the vacuum and satisfies the same 
anti-commutation relations as $a_{k\sigma r}$. Indeed, one has
e.g.,
\begin{eqnarray}
&&[\beta_{k\sigma r},\beta_{k'\sigma' r'}^\dag]_+
=
\widehat \Omega a_{k\sigma r}\widehat \Omega^\dag \widehat \Omega a_{k'\sigma' r'}^\dag \widehat \Omega^\dag
+\widehat \Omega a_{k'\sigma' r'}^\dag \widehat \Omega^\dag
\widehat \Omega a_{k\sigma r}\widehat \Omega^\dag 
=\widehat \Omega [a_{k\sigma r},a_{k'\sigma' r'}^\dag]_+
\widehat \Omega^\dag \widehat \Omega \widehat \Omega^\dag \cr
&&~~=\delta(k-k')\delta_{\sigma \sigma'} \delta_{rr'}
\widehat \Omega \widehat \Omega^\dag \widehat \Omega \widehat \Omega^\dag
=\delta(k-k')\delta_{\sigma \sigma'} \delta_{rr'} \ .
\nonumber
\end{eqnarray}
Also the intertwining relation $H_F\widehat \Omega=\widehat \Omega H_0$ shows that they are
normal modes of $H_F$:\break
$[H_{F},\beta_{k\sigma r}]=\widehat \Omega[H_0,a_{k\sigma r}]\widehat \Omega^\dag
=-\omega_{kr} \widehat \Omega a_{k\sigma r}\widehat \Omega^\dag
=-\omega_{kr} \beta_{k\sigma r}$.
The NESS averages of a product of `in' fields are then calculated as
follows:
\begin{eqnarray}
&&\langle \beta_{k_1\sigma_1 r_1}^\dag\beta_{k_2\sigma_2 r_2}^\dag 
\beta_{k_3\sigma_3 r_3}
\beta_{k_4\sigma_4 r_4}\rangle_{ss}
={\langle \widehat \Omega^\dag \beta_{k_1\sigma_1 r_1}^\dag
\beta_{k_2\sigma_2 r_2}^\dag \beta_{k_3\sigma_3 r_3}
\beta_{k_4\sigma_4 r_4}\widehat \Omega\rangle_0/\langle \widehat \Pi\rangle_0} \cr
&&~~={\langle \widehat \Omega^\dag \widehat \Omega a_{k_1\sigma_1 r_1}^\dag \widehat \Omega^\dag
\widehat \Omega a_{k_2\sigma_2 r_2}^\dag \widehat \Omega^\dag \widehat \Omega 
a_{k_3\sigma_3 r_3} \widehat \Omega^\dag \widehat \Omega a_{k_4\sigma_4 r_4}\widehat \Omega^\dag 
\widehat \Omega\rangle_0/\langle \widehat \Pi\rangle_0}
\cr
&&~~=
{\langle \widehat \Pi a_{k_1\sigma_1 r_1}^\dag \widehat \Pi a_{k_2\sigma_2 r_2}^\dag 
\widehat \Pi a_{k_3\sigma_3 r_3}\widehat \Pi 
a_{k_4\sigma_4 r_4}\widehat \Pi  \rangle_0
/ \langle \widehat \Pi \rangle_0}
\cr
&&~~=
{\langle a_{k_1\sigma_1 r_1}^\dag a_{k_2\sigma_2 r_2}^\dag 
a_{k_3\sigma_3 r_3} 
a_{k_4\sigma_4 r_4}\widehat \Pi^4  \rangle_0
/ \langle \widehat \Pi \rangle_0}
\cr
&&~~=
{\langle a_{k_1\sigma_1 r_1}^\dag a_{k_2\sigma_2 r_2}^\dag 
a_{k_3\sigma_3 r_3} 
a_{k_4\sigma_4 r_4}\rangle_0 \langle \widehat \Pi  \rangle_0
/ \langle \widehat \Pi \rangle_0}
=\langle a_{k_1\sigma_1 r_1}^\dag a_{k_2\sigma_2 r_2}^\dag 
a_{k_3\sigma_3 r_3} a_{k_4\sigma_4 r_4} \rangle_0 \ , \nonumber
\end{eqnarray}
where we have used $[a_{k\sigma r},\widehat \Pi]=0$, $\widehat \Pi^4=\widehat \Pi$
and the fact that $a_{k\sigma r}$ and $f_\sigma$ (hence $\widehat \Pi$) are 
independent in the initial state $\rho_l$.
Repeating the same arguments and reminding that the initial state 
$\rho_l$ satisfies Wick's theorem for $a_{k\sigma r}$, one finds that
the NESS satisfies Wick's theorem for the `in' fields 
$\beta_{k\sigma r}$ and, in particular, the two-point functions are
given by
\begin{equation}
\langle \beta _{k\sigma r}^{\dagger } \beta _{k'\sigma ' r'}\rangle_{ss} 
= F_r(\omega_{kr})\delta (k-k'
)\delta _{\sigma \sigma '}\delta _{rr'}.  \label{soukan}
\end{equation}
where $F_r(\epsilon)=1/(e^{\beta_r(\epsilon-\mu_r)}+1)$ stands
for the Fermi distribution function of each electrode.
Reminding that the `in' states are free particle states at $t=-\infty$, eq.(\ref{soukan}) 
simply implies that the particles originally coming 
from the left (right) electrode carries its initial statistical 
information such as the temperature and chemical potential.

In order to close the self-consistent equations, one task is left, namely
the construction of the `in' fields. As shown in Appendix, they are obtained
from an operator equation $[H_F,\beta_{k\sigma r}]=-\omega_{kr} \beta_{k\sigma r}$ with the boundary 
condition: $a_{k\sigma r}(t) \exp(i\omega _{kr}t/\hbar ) \rightarrow 
\beta_{k\sigma r}$ as $t\rightarrow -\infty $.
The so-obtained relations can be inverted and the original operators are
expressed in terms of the `in' fields:
\begin{eqnarray}
&&f_\sigma =\sum_{r=L,R} \int dk z_\sigma u_{kr}{\chi_\sigma(\omega_{kr})^*
\beta_{k\sigma r}\over \Lambda_\sigma(\omega_{kr})^*} \label{original1}
\\
&&a_{k\sigma r}=\beta _{k\sigma r}
+\int dk'\left(\dfrac{z_{\sigma }^2u_{kr}u_{k'r}\beta_{k'\sigma r}}
{\omega _{k'r}-\omega _{kr}+i0}
\dfrac{\xi _{\sigma }(\omega _{k'r})^{*}}{\Lambda_{\sigma }(\omega_{k'r})^{*}}
+\dfrac{z_{\sigma }^2 u_{kr}u_{k'\bar r}\beta _{k'\sigma \bar r}}{\omega _{k'\bar r}-
\omega _{kr}+i0}\dfrac{\kappa _{\sigma }(\omega _{k'\bar r})^{*}
}{\Lambda_{\sigma }(\omega_{k'\bar r})^{*}}\right) 
\label{original2}
\end{eqnarray}
where $r=L$ or $R$ and we abbreviate $\bar L=R$, $\bar R=L$.
The auxiliary functions are defined as
\begin{eqnarray}
\chi _{\sigma }(x)&=&1+We^{i\varphi _{r}}\Omega _{\sigma r}(x)/z_{\sigma }\cr
\Lambda_{\sigma }(x)&=&\nu_{\sigma }(x) (x-\bar \epsilon _{\sigma })-\sum_r 
\Omega _{\sigma r}(x)-2W\cos\varphi \Omega _{\sigma L}(x)
\Omega _{\sigma R}(x)/z_{\sigma }\cr
\nu_{\sigma }(x)&=&1-W^2\Omega_{\sigma L}(x)
\Omega _{\sigma R}(x)/z_{\sigma }^2 \cr
\xi _{\sigma }(x)&=&1+\Omega_{\sigma r}(x)
[W^2(x-\bar \epsilon_{\sigma })/z_{\sigma}^2+2W
\cos\varphi /z_{\sigma }]\cr
\kappa _{\sigma }(x)&=&1+We^{i\varphi_{r}}(x-\bar \epsilon_{\sigma })/
z_{\sigma }\nonumber
\end{eqnarray}
where $\Omega_{\sigma r}(x)=\displaystyle \int dk''
\frac{u_{k'' r}^2z_{\sigma }^2}{x-\omega _{k'' r }-i0}$
and $\varphi_L$ ($\varphi_{R}$) stands for $\varphi$ ($-\varphi$).

By combining (\ref{original1}), (\ref{original2}) with (\ref{soukan}), 
one can calculate
$\langle f_\sigma^\dag f_\sigma\rangle_{ss}$ and $M_\sigma$ in 
(\ref{WeakerConstraint1}),
(\ref{WeakerConstraint2}), (\ref{SelfST1}),
(\ref{SelfST2}) and (\ref{SelfST3})
(or in 
(\ref{selfconsistent1}), (\ref{selfconsistent2}) and (\ref{selfconsistent})). 
This completes 
the derivation of the self-consistent equations.
We note that, for equilibrium states, the self-consistent 
equations derived here reduce to those obtained by 
Kotliar and Ruckenstein~\cite{Kotliar} via a saddle-point approximation.

\section{Basic Equations}\label{Sec3}
From now on, we assume that the wave-number dependence of the tunneling
matrix elements is weak so that $u_{kr}\equiv u_r$ ($r=L,R$) is constant 
and that the electrode-dot coupling is symmetric: 
$2\pi u_L^2\rho_L
=2\pi u_R^2\rho_R\equiv \Gamma$, where
$\rho_{L/R}$ is the density of states of the left/right electrode.
Moreover, we neglect the Zeeman energy and 
limit our attention to the spin degenerate case.
Then, the mean fields, the average electron number per spin at the dot as well as
the original dot-level are independent of the spin
quantum number: $z_{\uparrow }=~z_{\downarrow }\equiv~z$,
$\bar \lambda _{\uparrow }^{(1)}=
\bar \lambda _{\downarrow }^{(1)}\equiv\lambda$, 
$\langle f_{\uparrow }^\dag f_{\uparrow }\rangle_{ss}
=\langle f_{\downarrow }^\dag f_{\downarrow }\rangle_{ss}\equiv~n$,
and
$\epsilon _{\uparrow }=\epsilon _{\downarrow }\equiv \epsilon _{0}$.
And, the self-consistent equations 
(\ref{selfconsistent1}), (\ref{selfconsistent2}) and 
(\ref{selfconsistent}) read as
\begin{align}
&z=\frac{\sqrt{(n-D)(1+D-2n)}+\sqrt{D(n-D)}}{\sqrt{n(1-n)}} 
\label{BasicSC1}\\
&\lambda =\frac{M}{\sqrt{n(1-n)}} \left[ 2\sqrt{\frac{n-D}{1+D-2n}}-\sqrt{\frac{1+D-2n}{n-D}}
-\sqrt{\frac{D}{n-D}}+\frac{z(1-2n)}{\sqrt{n(1-n)}} \right]
\label{BasicSC2}\\
&4\frac{dz}{dD}M+U=0
\label{BasicSC3}
\end{align}
where
\begin{equation}
n=\frac{\Gamma z^2}{2\pi }(1+z^4x)\int d\epsilon \frac{1}{|\Lambda (\epsilon )|^2}(F_{L}(\epsilon )+F_{R}(\epsilon ))
+\frac{\Gamma z^4}{\pi }\sqrt{x}\sin \varphi \int d\epsilon \frac{1}{|\Lambda (\epsilon )|^2}(F_{L}(\epsilon )-F_{R}(\epsilon )) \label{BasicSC4} \\
\end{equation}
\begin{align}
M=\frac{\Gamma z}{2\pi }(1+z^2x) \int d\epsilon &\frac{(\epsilon -\epsilon _{0}+\lambda )}{|\Lambda (\epsilon )|^2}(F_{L}(\epsilon )+F_{R}(\epsilon )) \notag \\
&+\frac{\Gamma z^2}{\pi }\sqrt {x}\sin \varphi \int d\epsilon \frac{(\epsilon -\epsilon _{0}+\lambda )}{|\Lambda (\epsilon )|^2}(F_{L}(\epsilon )-F_{R}(\epsilon ))\label{BasicSC5}
\end{align}
\noindent
with $\lvert \Lambda (\epsilon ) \rvert^2
=[(1+x)(\epsilon /\Gamma - (\epsilon_{0}-\lambda)/\Gamma )+\sqrt x\;  z^2\cos \varphi ]^2
+z^4$ and $x\equiv \pi ^2W^2u_L^2 \rho _{L} \mskip3mu u_R^2\rho _{R}$ measures the strength
of the transmission via the direct tunneling. 
The chemical potentials $\mu_r$ of the electrodes, appeared in
the Fermi distribution functions
$F_r(\epsilon)=1/(e^{\beta_r(\epsilon-\mu_r)}+1)$  ($r=L,R$), 
are taken to be $\mu_L=\mu-eV/2$, $\mu_R=\mu+eV/2$, where $\mu$ is the
average chemical potential.
The probability $D$ of the double occupation at the dot level and
the Lagrange multiplier $\lambda$ are obtained as 
solutions of these self-consistent equations.

Now we turn to the average current at the nonequilibrium steady state.
The current operator from left to right electrodes is expressed by the 
time derivative 
of the number operator of the left electrode 
$N_L=\sum_\sigma \int dk a_{k\sigma L}^{\dagger}a_{k\sigma L}$:
\begin{eqnarray}
{\hat J}&\equiv& -e {d\over dt}N_L={ie\over \hbar}[N_L,{\widetilde H}]\cr
&=&\frac{ie}{\hbar}\sum_\sigma \left[\int dku_{kL} {\hat z}_\sigma a_{k\sigma L}^{\dagger } f_{\sigma }
+ \iint dkdqWu_{kL}u_{qR}
e^{i\varphi }
a_{k\sigma L}^{\dagger }a_{q\sigma R}
-({\rm H.C.})
\right] \ .
\end{eqnarray}
Its NESS average is given by
\begin{eqnarray}
\langle \hat J \rangle_{ss} 
&=&\frac{ie}{\hbar}\sum_\sigma \left[\int dku_{kL} z_\sigma \langle a_{k\sigma L}^{\dagger } f_{\sigma }\rangle_{ss}
+ \iint dkdqWu_{kL}u_{qR}
e^{i\varphi }
\langle a_{k\sigma L}^{\dagger }a_{q\sigma R}\rangle_{ss}
-({\rm c.c.})
\right] \cr
&=&
-\dfrac{2e}{h}\; \displaystyle \int_{\mu_c}^{\infty}d\epsilon \; T(\epsilon )
\{F_L(\epsilon )-F_R(\epsilon )\}. \label{eq:current}
\end{eqnarray}  
where the first term is evaluated as 
$\langle {\hat z}_\sigma a_{k\sigma L}^{\dagger } f_{\sigma }\rangle_{ss}
= {z}_\sigma \langle a_{k\sigma L}^{\dagger } f_{\sigma }\rangle_{ss}$ 
and (\ref{soukan}) is used.
In this equation, $T(\epsilon)$ is the transmission probability: 
\begin{equation}
T(\epsilon )=\frac{z^4}{\lvert \Lambda(\epsilon ) \rvert ^2}\left\{ \left[ \frac{2\sqrt x}
{z^2}\left(\frac{\epsilon }{\Gamma }
-\frac{\epsilon _{0}-\lambda}{\Gamma }\right)+\cos \varphi \right]^2
+\sin ^2 \varphi \right\}
\ .
\end{equation}  

It is interesting to rewrite the transmission probability 
in the Fano form \cite{NRG,Bjoen,Ueda}:
\begin{equation}
T(\epsilon )=T_{b}\frac{|\tilde E+q|^2}{\tilde E^2+1}
\label{FanoForm}
\end{equation}
where $\tilde E=~(1+~x)\{(\epsilon -\epsilon_{0}+\lambda)/(z^2\Gamma )+[\sqrt x /(1+x)]\cos \varphi \}$ 
and $T_{b}=4x/(1+x)^2$ is the background transmission probability \cite{NRG}.
The Fano parameter $q=[(1-x)\cos \varphi +i(1+x)\sin \varphi ]/2\sqrt x$ 
takes into account the interference effect and determines
the peak shape of the conductance. 
The conductance shows the Breit-Wigner type resonance if $x=0$ 
(the real part of $q$ is $\infty$)
and the anti-resonance if $x=1$ (the real part of $q$ is $0$).   
Eq. (\ref{FanoForm}) also indicates that the 
SBMF approach takes into account the effect of Coulomb interaction 
in two ways, namely as renormalization of the dot energy by $\lambda$:
$\epsilon_0\to \epsilon_0-\lambda$, and 
as reduction of the electrode-dot interaction by a factor
of $z^2$: $\Gamma \to z^2 \Gamma$. 
The energy renormalization is partly taken into account in a simpler 
Hartree approximation. The reduction of the electrode-dot interaction
implies the increase of the lifetime of the dot level and is interpreted
as the stabilization of the dot level due to strong correlation between
the conduction and localized electrons, which might be identified
with precursor of the formation of a singlet quasi-bound state between the two and, thus,
with development of the Kondo correlation.

In the following sections, we investigate transport properties
at zero temperature,\break
$1/\beta_L=1/\beta_R=0$, in detail.

\section{Transports in Linear Response Regime}\label{Sec4}

In this section, we focus on the linear response regime where the 
conductance is given by
$$
G={2e^2\over h}T(\mu)
$$
and the mean fields $D$ and $\lambda$ are determined by 
(\ref{BasicSC1}), (\ref{BasicSC2}), (\ref{BasicSC3}), (\ref{BasicSC4})
and (\ref{BasicSC5})
with\break
$V=0$.
Then, we
compare the results with those by the numerical renormalization group
(NRG) technique\cite{NRG}.

\subsection{Fano-Kondo resonances}
\begin{figure}[h]
\begin{center}
\includegraphics[width=9cm]{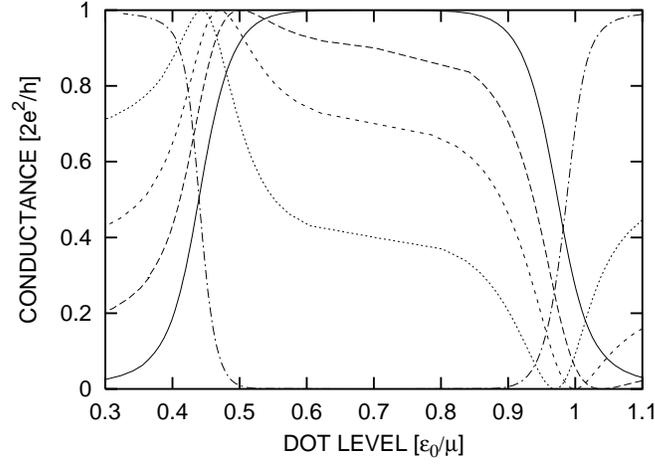}
\end{center}
\caption{The conductance in the linear-response regime as a function of the original dot level $\epsilon _{0}/\mu$, 
where $\varphi=0$, $\Gamma /\mu =0.02$ and the Coulomb interaction is $U/\mu=0.6$ ($U/\Gamma =30$) with $\mu$ the Fermi energy. 
Solid curve is the background transmission probability $T_{b}=0.0$, 
the long-dashed curve at $T_{b}=0.1$, the short-dashed curve at $T_{b}=0.3$, the dotted curve at $T_{b}=0.6$
and the dash-dotted curve at $T_{b}=1.0$.}
\label{conductance}
\end{figure}

\begin{figure}[b]
  \begin{center}
 \begin{tabular}{cc}
     \resizebox{70mm}{!}{\includegraphics{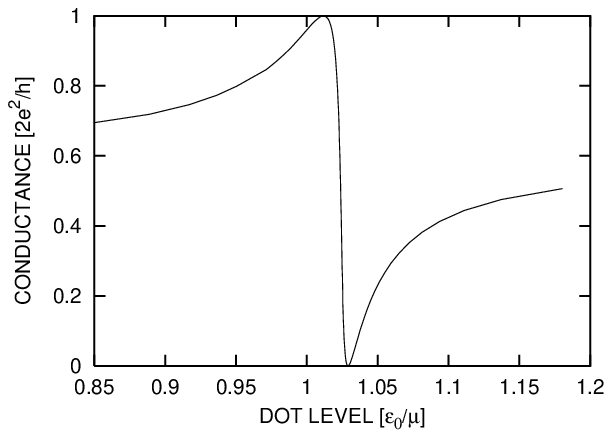}} &
      \resizebox{70mm}{!}{\includegraphics{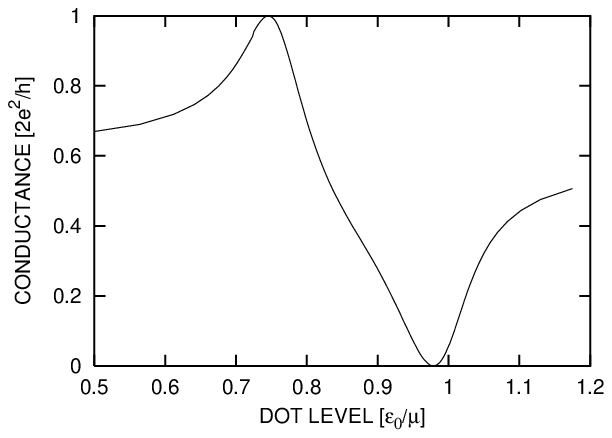}} \\
      \resizebox{70mm}{!}{\includegraphics{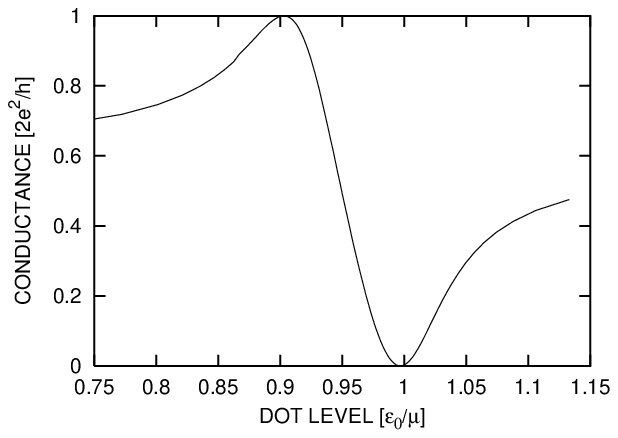}} &
      \resizebox{70mm}{!}{\includegraphics{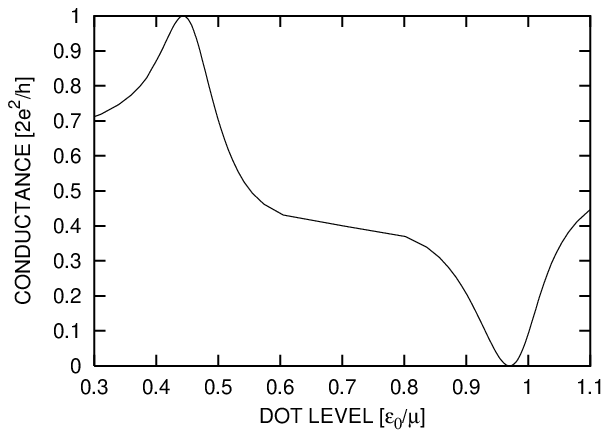}} \\
    \end{tabular}
    \caption{The Coulomb interaction dependence of the conductance with $\varphi=0.0, T_{b}=0.6$. 
The Fano-Kondo plateau is formed with increasing the Coulomb interaction,
the left side of the upper panel is for $U/\mu=0.02$, the right side for $U/\mu=0.3$ and 
the left side of the lower panel for $U/\mu=0.14$, the right side for $U/\mu=0.6$.}
    \label{CoulombFano}
  \end{center}
\end{figure}

Figure \ref{conductance} shows the conductance as a function of
the original dot level $\epsilon_{0}$ 
for various values of the background transmission probability $T_b$
where the phase difference
due to the magnetic flux, the electrode-dot coupling
and the Coulomb energy
are taken, respectively, as
$\varphi=0$, $\Gamma/\mu=0.02$ and $U/\mu=0.6$
with $\mu$ the average chemical potential.
When $T_{b}=0.0$, the conductance reaches the unitary limit $G=2e^2/h$
and the well-known Kondo plateau is formed.
With increasing the background transmission probability 
(cf. the curves for $T_{b}=0.1, T_{b}=0.3$), the Kondo plateau
is modified by the Fano resonance due to the direct channel
transport.
When $T_b=0.6$, the conductance curve has an asymmetric form, typically
observed in the Fano resonance, with a large plateau between the peak
and dip. The plateau is generated by the strong Coulomb interaction.
Indeed, as shown in Fig. \ref{CoulombFano}, the plateau region 
is gradually formed by increasing the Coulomb interaction.
When the transport through the direct channel is dominant (i.e., 
$T_{b}=1.0$), the destructive interference causes the anti-resonance.
Such a behavior is experimentally
observed in the T-shaped QD system \cite{Katsumoto,Sato}.

We would like to remark that all the curves in Fig.\ref{conductance} 
are in good agreement
with those in Fig.2 of Ref.\citen{NRG}, that is obtained by the 
NRG technique.
The parameters of the two results are almost the same except the strength 
of the Coulomb interaction $U/\Gamma\simeq 30$, which is almost twice 
larger than that of Ref.\citen{NRG}.
This indicates that the essential features of the conductance such as 
the shapes or peak positions are well captured by the SBMF approach 
in spite of neglecting the fluctuations of the bosonic fields and
Lagrange multipliers.

\subsection{Conductance plateau}

\begin{figure}[tb]
  \begin{center}
 \begin{tabular}{cc}
       \resizebox{70mm}{!}{\includegraphics{conductanceTb=0.6-1.eps}} &
       \resizebox{70mm}{!}{\includegraphics{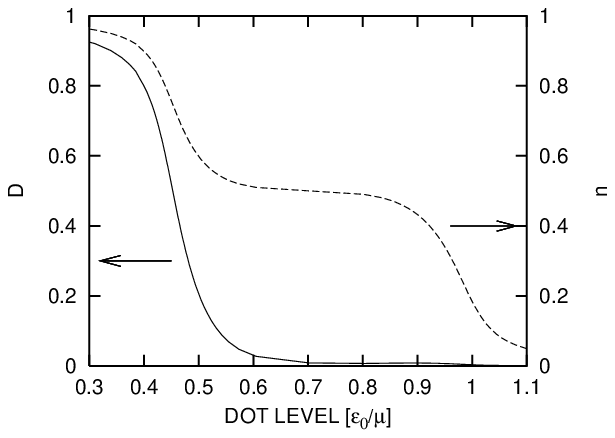}} \\
      \resizebox{70mm}{!}{\includegraphics{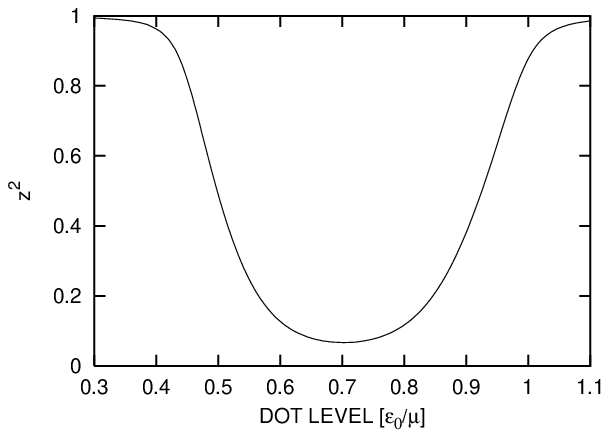}} &
      \resizebox{70mm}{!}{\includegraphics{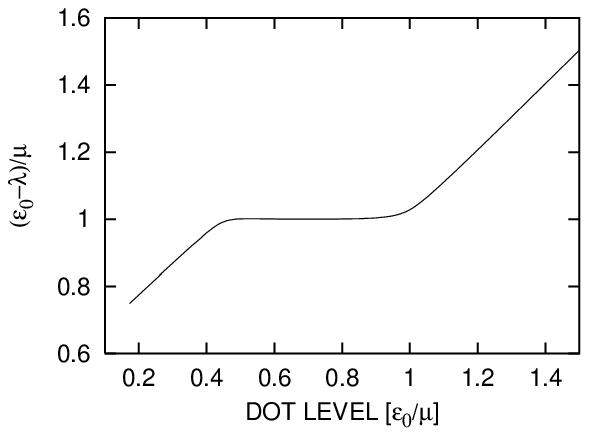}} \\
    \end{tabular}
    \caption{SBMF parameters as functions of the dot energy. 
The left side of the upper panel is the differential conductance with $T_{b}=0.6$, 
the right side is the electron occupation number and double occupation probability,
the left side of the lower panel is the reduction factor of the dot level-width $z^2$, 
and the right side is the renormalized dot energy as functions of the original dot level.
The Coulomb interaction and AB phase are set to be $U/\Gamma =0.6$, $\varphi =0$.}
    \label{KRparameter}
  \end{center}
\end{figure}

Figure \ref{CoulombFano} indicates that strong Coulomb interaction is 
responsible for the plateau formation in the conductance curve. 
The underlying physical mechanism becomes more transparent by studying
the average number of electrons per spin occupying the dot: $n$, and 
the probability of simultaneously finding two electrons at the dot: $D$.
The left side of the upper panel in Fig.\ref{KRparameter} shows the 
conductance at $T_{b}=0.6$ and 
the right side shows the average occupation number $n$, and 
the double occupation probability $D$.
When the dot level is far above the Fermi energy 
($\epsilon_0/\mu \gtrsim 1.0$), 
no electrons are found in the QD.
As the dot level lowers, the occupation number $n$ increases and 
reaches $1/2$ at $\epsilon_{0}/\mu \simeq 0.9$.
Within the interval $0.5 \lesssim \epsilon _{0}/\mu \lesssim 0.9$, one sees
$n\simeq 1/2$ and $D\simeq 0$, which implies that 
the dot is singly occupied and
the second electron can not enter 
into the dot owing to the strong Coulomb interaction.
In this interval, the electronic state does not change and, as a result, 
the conductance shows the wide plateau.
When the dot level lowers further, the second electron starts to flow into the
dot as indicated by the increase of the double occupation probability $D$,
and the conductance again starts to form a peak.

It is interesting to see the associated behavior of
the SBMF parameters.
The lower panels in Fig.\ref{KRparameter} show the reduction 
factor $z^2$ of the dot level-width and  
the renormalized dot energy $\epsilon_0-\lambda$.
Outside the plateau region $0.5 \lesssim \epsilon _{0}/\mu \lesssim 0.9$, 
$z^2\simeq 1$ and $\epsilon_0-\lambda \propto \epsilon_0+({\rm const.})$.
Hence, the dot level-width and the renormalized dot energy
are the same as their bare values and the behavior of the system is 
essentially equivalent to that of the noninteracting case. 
On the other hand, within the plateau region 
$0.5 \lesssim \epsilon _{0}/\mu \lesssim 0.9$, 
$\epsilon_0-\lambda$ is almost constant and $z^2$ is very small. 
The former indicates that the Coulomb interaction modifies the effective 
dot energy seen by the reservoir electrons so that the dot level is kept 
to be singly occupied.
The latter indicates that the dot level-width is
drastically suppressed by the Coulomb repulsion. 
The formation of the plateau is mainly due to the 
renormalization of the dot energy. 
As will be discussed in the next section, the effects of the reduction
of the dot level-width can be seen more clearly in nonequilibrium
regime.

\subsection{Phase dependence of conductance}
\begin{figure}[tb]
\begin{center}
 \includegraphics[width=9cm]{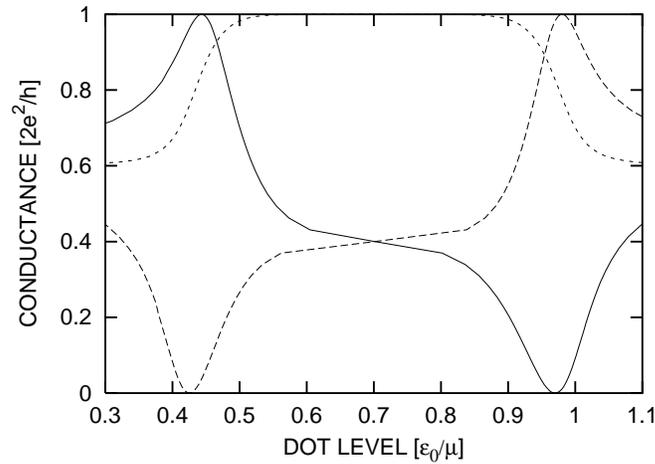}
\end{center}
\caption{The conductance for various values of the AB phase: $\varphi=0$ (solid curve), $\varphi=\pi/2$ (short-dashed curve), 
$\varphi=\pi$ (long-dashed curve). The other parameters are set to be $T_{b}=0.6$, $U/\mu=0.6$. }
\label{phaseconductance}
\end{figure}

Figure \ref{phaseconductance} illustrates the phase dependence 
of the Fano-Kondo resonance.
As in the noninteracting case, with changing the AB phase from 
$\varphi=0$ to $\varphi=\pi$, 
the destructive interference between the localized and continuous states
turns into the constructive interference and vice versa.
On the other hand, the plateau of the two curves are almost the same. 
At $\varphi=\pi/2$,
the conductance has a symmetric resonance peak and reaches the unitary 
limit value as in the case of $T_{b}=0.0$ and $\varphi=0$ (cf. Fig.\ref{conductance}).

\begin{figure}[tb]
\begin{center}
\includegraphics[width=9cm]{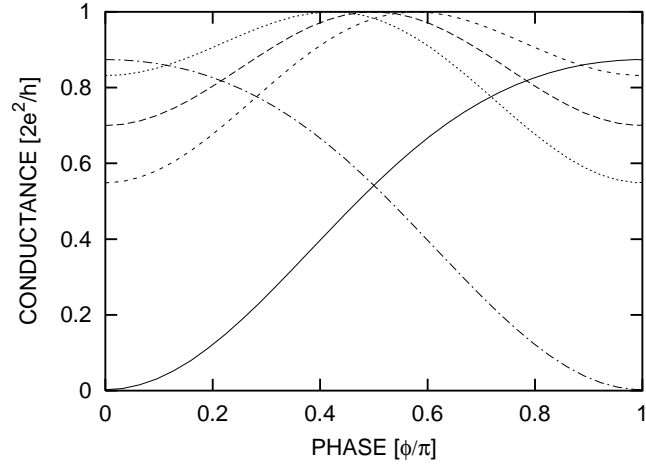}
\end{center}
\caption{AB oscillations as a function of the AB phase for the fixed electron occupation number with
$T_{b}=0.3$ and $U/\mu=0.6$. 
Individual curves correspond to $n=0.2$ (solid curve), $n=0.45$ (short-dashed curve), $n=0.5$ (long-dashed curve),
$n=0.55$ (dotted curve) and $n=0.8$ (dash-dotted curve). }
\label{ABoscillation}
\end{figure}

We now turn to the relation of the AB oscillations with the 
Fano-Kondo plateau.
It was shown that
the phase of the AB oscillation at one side 
of the conductance peak is different by $\pi$ from that 
at the other side\cite{Yacoby,Schuster}. 
Also the strong Coulomb 
interaction is known to fix the maximum position of the conductance
at $\varphi\simeq\pi/2$ 
in the middle of the Fano-Kondo plateau \cite{ABphase}.
Figure \ref{ABoscillation} shows the conductance 
as a function of the AB phase $\varphi$ for various occupation numbers $n$ of the
dot level.
Outside the plateau regime ($n=0.2$, 0.8), 
the conductance 
shows a $2\pi$-periodic AB 
oscillation and dose not reach the unitary-limit value $2e^2/h$.
In contrast, in the plateau regime ($n=$ 0.45,0.5,0.55), the 
conductance reaches the unitary limit $2e^2/h$ near 
$\varphi=\pi/2$ and is similar to a $\pi$-periodic sinusoidal curve.
In other words, in the plateau regime, the period of the AB oscillation
is almost fixed to be $\pi$ and the conductance takes its maximum 
near $\varphi=\pi/2$.
This behavior is different from the noninteracting case where the AB
oscillation gradually changes its character as the increase of $n$\cite{JPSJ1}.
Note that these features of the AB oscillation in the interacting 
case are consistent with the experiments and theoretical predictions of 
the works mentioned above\cite{Katsumoto,Aikawa,NRG,ABphase}.

\section{Transports in Nonequilibrium Regime}\label{Sec5}

It is often argued that SBMF approaches are not appropriate for 
studying the dynamical properties of strongly correlated systems since
they can not deal with the effect of charge 
fluctuations\cite{Ramon,Dong2}.
We have shown that the finite-$U$ SBMF 
approach describes well, at least qualitatively, the transports 
of the AB ring with a QD in the linear response regime as compared 
with the NRG approach\cite{NRG}. 
So we believe that the SBMF approach could catch the essential 
features of the steady-state transports at finite bias voltage $V$.
In this section, 
we study the properties of the differential conductance at fixed
occupation number $n$:
$$
G_n=\left({\partial \langle \hat J\rangle 
\over \partial V}\right)_n \ ,
$$
where the subscript $n$ means that the $V$-derivative is taken by keeping 
$n$ constant. 
Instead of the differential conductance at fixed dot level $\epsilon_0$,
here we consider $G_n$ because of a technical difficulty in solving 
the self-consistent equations.  
But the difference is small enough for describing
the essential features except
the region of $|hG_n/(2e^2)|\ll 1$. For example, 
at $\varphi=0$, $\Gamma/\mu=0.02$,
$U/\mu=0.6$, $n=0.43$ and $eV/\mu=0.002$,
the difference is estimated to be at most $1\%$. 

\begin{figure}[b]
  \begin{center}
 \begin{tabular}{cc}
    \resizebox{70mm}{!}{\includegraphics{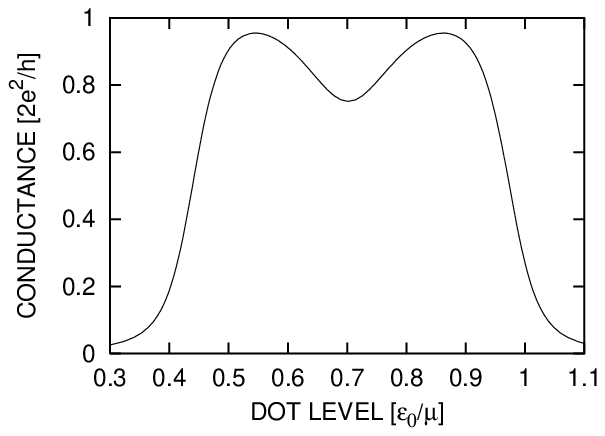}} &
      \resizebox{70mm}{!}{\includegraphics{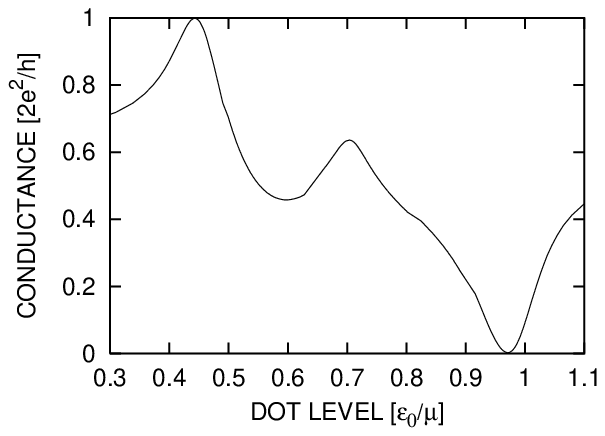}} \\
      \resizebox{70mm}{!}{\includegraphics{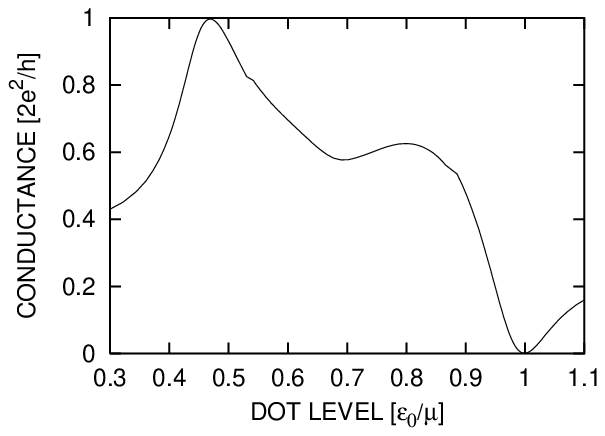}} 
    \end{tabular}
    \caption{Differential conductance at nonequilibrium regime as a function of the original dot level
for several background transmission probabilities $T_{b}=0.0$, $T_{b}=0.6$ (left and right of upper panel)
and $T_{b}=0.3$ (left side of lower panel) with $\varphi =0$, $U/\mu=0.6$ and $eV/\mu=0.002$.}
    \label{NonequilibriumFano}
  \end{center}
\end{figure}

Figure \ref{NonequilibriumFano} shows the $\epsilon_0$-dependence of 
$G_n$ at $eV/\mu=0.002$
for various values of the 
background transmission $T_{b}$ with $\varphi=0$, $\Gamma/\mu=0.02$ 
and $U/\mu=0.6$.
At $T_{b}=0.0$, 
the conductance plateau is suppressed by the low bias voltage.
When $T_{b}=0.3$, the bias voltage induces a peak in the conductance
plateau. 
Such a behavior is recently observed in the 
side-coupled QD \cite{Katsumoto}, which is one of the systems 
exhibiting the Fano resonance.
As the background transmission increases, the resonance peak of
the differential conductance tends to split into two asymmetric peaks
(cf. the case of $T_{b}=0.6$). 

This could be understood as follows.
Recall that the similar splitting of the resonance peak has been
observed in the noninteracting case at higher bias voltage.
In this case, a conductance peak is formed once the dot
energy passes the Fermi level of the electrodes.
At high bias voltage, two electrodes have very different Fermi 
energies and the two peaks are observed. While, if the bias voltage
is so low that the difference between the Fermi levels is less than
the width of the dot level, the two peaks are degenerate and one
observes only one peak.
On the other hand, as mentioned after (\ref{FanoForm}), the transmission
probability in the interacting case has the similar form as in the
noninteracting case except the renormalization of the dot energy and the 
reduction of the dot level-width. Therefore, if the bias voltage is
large enough compared with the level-width, two peaks can be observed. In the plateau regime, 
the reduction factor $z^2$ is small (cf. Fig. \ref{KRparameter}) 
and there appear two peaks even
at very low bias voltage. In other words, because of the reduction 
of the width, the system becomes more sensitive to the bias voltage
and the splitting of the resonance peak is observed at low bias 
voltage.

\begin{figure}[tb]

\begin{center}
\includegraphics[width=9cm]{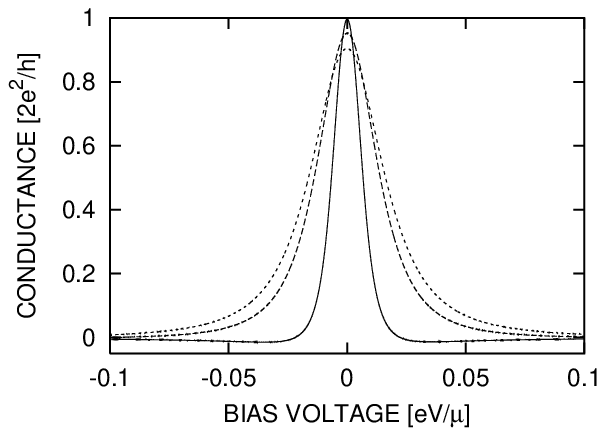}
\end{center}
\caption{The differential conductance as a function of the bias voltage with $T_{b}=0.0$, $\varphi=0$ and 
$U/\mu=0.6$ for several fixed occupation numbers: $n=0.48$ (solid curve), 
$n=0.43$ (long-dashed curve) and $n=0.40$ (short-dashed curve). At $n=0.48$ and $eV/\mu =0$, the conductance reaches the unitary limit.}
\label{bias1}
\end{figure}

\begin{figure}[tb]

\begin{center}
\includegraphics[width=9cm]{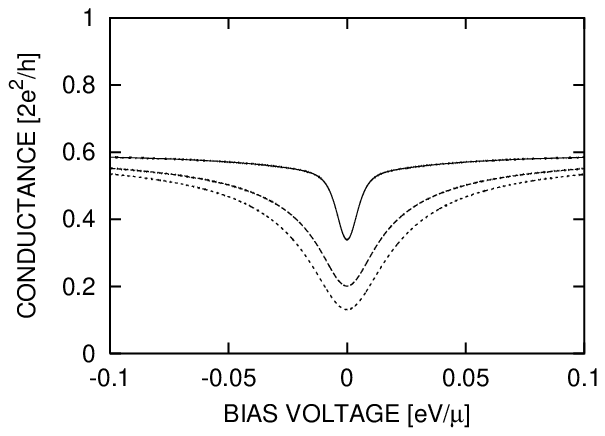}
\end{center}
\caption{The differential conductance as a function of the bias voltage with $T_{b}=0.6$, $\varphi=0$ and 
$U/\mu=0.6$ for several fixed occupation numbers: $n=0.48$ (solid curve), 
$n=0.43$ (long-dashed curve) and $n=0.40$ (short-dashed curve). }
\label{bias2}
\end{figure}

\begin{figure}[tb]
\begin{center}
 \includegraphics[width=9cm]{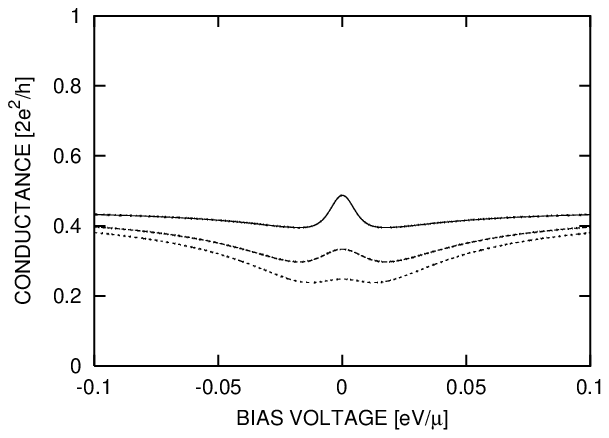}
\end{center}
\caption{The differential conductance as a function of the bias voltage with $T_{b}=0.45$, $\varphi=0$ and 
$U/\mu=0.6$ for several fixed occupation numbers: $n=0.48$ (solid curve), 
$n=0.43$ (long-dashed curve) and $n=0.40$ (short-dashed curve). }
\label{bias3}
\end{figure}

The differential conductance $G_n$ is shown as a function of the bias voltage 
in Fig.\ref{bias1}, Fig.\ref{bias2} and Fig.\ref{bias3} 
for various occupation numbers, where the other parameters are set again
as $\varphi=0$, $\Gamma/\mu=0.02$ and $U/\mu=0.6$. 
When the background conduction is absent $T_{b}=0.0$ (i.e., in case of the transport through a single QD),
the conductance has a sharp peak at zero bias voltage. These peaks correspond to 
those experimentally observed in refs.\citen{Kondoeffect1,Kondoeffect2}. 
With decreasing the occupation number (this corresponds to the increase of $\epsilon_0$
from the center of the conductance plateau of Fig.\ref{conductance}),
the zero-bias maximum is reduced from the unitary limit value $2e^2/h$ and the width of the 
peak becomes broader  (cf. the curves at $n=0.48$, 0.43, 0.40).
This behavior of the zero-bias maximum agrees with the calculations
reported by Craco {\it et al. }\cite{Craco} and by B. Dong {\it et al. }\cite{Dong2,fnote3}.
Note that $G_n$ at $n=0.48$ has very small negative value for larger $|V|$, which
seems to be an artifact due to the use of $G_n$.

In contrast, when the direct tunneling is turned on, the conductance chnages its behavior.
At $T_{b}=0.6$ (Fig.\ref{bias2}), there appears a zero bias dip and, similarly to the zero bias
peak at $T_{b}=0.0$, the dip width becomes broader as $n$ decreases.
Moreover, at $T_{b}=0.45$ (Fig.\ref{bias3}), a peak gradually changes into a broad dip as the decrease of $n$.
In case of infinite Coulomb repulsion $U\to +\infty$, Bu{\l}ka {\it et. al.} \cite{Bulka} predicts
the possibility of the zero bias dip depending on the magnetic flux piercing the AB ring.
Here we have shown that the similar behavior is possible depending on the strength of 
the background transmission.

\section{Conclusions}\label{Sec6}
We have investigated the transports through the QD embedded in 
the AB ring in the linear response
and finite bias regimes, employing the nonequilibrium 
finite-$U$ SBMF approach.
The mean fields are evaluated at a nonequilibrium 
steady state with respect to the mean field
Hamiltonian which is constructed with the aid of the C$^*$-algebraic 
approaches.
The Coulomb interaction is found to be taken into account as the renormalization of 
the dot
energy and the reduction of the dot level-width. In particular, the 
latter may be interpreted as
a signature of the formation of the singlet bound state between the dot 
and conduction electrons.
We have studied the Fano-Kondo resonances and AB oscillations at the 
linear response regime
and obtained the results which, by using twice larger Coulomb energy, 
are in good agreement with those by the 
numerical renormalization group (NRG) technique.
Furthermore, this method is applied to the case with finite bias voltage, and 
we found that the Fano-Kondo plateau is suppressed and deformed by small 
bias voltage and that, in addition to the zero-bias peak, the zero-bias 
dip can be observed in the differential 
conductance depending on the strength of the background transmission. 

It is interesting to give a rough estimate of the Kondo temperature. 
If the SBMF parameters were 
weakly dependent on temperature, the linear-response conductance at 
$T_b=0$ and $\epsilon_0-\lambda=\mu$ 
reads as
$$
{h\over 2e^2}G=\int_{-\infty}^{+\infty}dy \left\{1+\left({Ty\over z^2\Gamma}\right)^2\right\}^{-1}{e^y\over (e^y+1)^2}
\simeq 1 - {1\over 3}\left({\pi T\over z^2\Gamma}\right)^2+{7\over 15}
\left({\pi T\over z^2\Gamma}\right)^4
+\cdots
\ ,
$$
for $T/(z^2\Gamma)\ll 1$. By comparing this with the low temperature expansion of the empirical 
formula\cite{HC}
$$
{h\over 2e^2}G=\left({(2^{1/s}-1)T_k^2\over T^2 + (2^{1/s}-1)T_k^2}\right)^s \ ,
$$
where $T_k$ stands for the Kondo temperature, one obtains $s=5/37$ 
and $T_k\simeq 2.64\times z^2\Gamma$. In case of $\Gamma/\mu=0.02$ and 
$U/\mu=0.6$, $z^2\simeq
0.066$ and $T_k\simeq 375$mK for $\mu=9.29$ meV. The value is 
consistent with Haldane's formula \cite{Haldane}:
$$
T_{k}=\frac{1}{2}\left(\frac{\Gamma U}{2}\right)^{1/2}\exp \left\{\frac{\pi\epsilon_{0}(U+\epsilon_{0})}{\Gamma U}\right\}
$$
which leads to $T_k\simeq 443$mK at $n=0.48$ as well as the 
experimental value of $T_k=450\sim 460$mK\cite{Kondoeffect1}. 
It should be noted that, without the reduction of the dot level-width, 
one would have a ten times larger value for the Kondo temperature. 
Moreover, this analysis shows that the reduction factor $z^2$ of the dot
level-width is essentially the Kondo temperature and, thus, the 
$\epsilon_0$-dependence of $z^2$ (cf. Fig.\ref{KRparameter}) can be 
reinterpreted as $\epsilon_0$-dependence of the Kondo temperature, 
which is parabolic near the center of the plateau and is consistent 
with experiments\cite{Kondoeffect1}.
These observations seem to support our interpretation that the reduction 
of the dot level-width implies the formation of the Kondo singlet.
However, to confirm this view, the finite temperature
case should be investigated and will be discussed elsewhere.

\section*{Acknowledgment}
The authors thank Professor M. Eto, Professor S. Katsumoto, Professor S. Pascazio and Professor T. Kato 
for fruitful discussions and valuable comments. 
One of them (JT) thanks Professor S. Katsumoto for private discussion 
for informing them of ref.\citen{Katsumoto} prior to publication. 
JT also thanks Professor T. Kato for information of ref.\citen{Aikawa} and
thanks Dr. H. Aikawa and Mr. M. Sato for discussing on their these.  
This work is partially supported by a Grant-in-Aid for Scientific 
Research of Priority Areas ``Control of Molecules in Intense Laser Fields'' 
(No. 14077219)
and the 21st Century COE Program at Waseda University 
``Holistic Research and Education Center for Physics of 
Self-organization Systems''
both from the Ministry of Education, Culture, Sports, Science and 
Technology of Japan, a Grant-in-Aid for Scientific Research (C) 
(No.17540365) from the Japan Society of the Promotion of Science
and Waseda University Grant for Special Research Projects (2004A-161,
research promotion 2005). 

{\appendix

\include{appa}
\section{Construction of `in' fields}

Here we derive the `in' fields and show (\ref{original1}) and 
(\ref{original2}).
Thanks to the bilinearity of $H_F$, the `in' fields $\beta_{k\sigma r}$ 
can be
written as
\begin{equation}
\beta_{k\sigma r}=a_{k\sigma r}+g_{r}^{(1)}(k) f_\sigma 
+ \sum_{r'}\int dk' g_{r: r'}^{(2)}(k,k') a_{k'\sigma r'} \ .
\end{equation}
Substituting this into $[H_F,\beta_{k\sigma r}]=-\omega_{kr} 
\beta_{k\sigma r}$, we obtain
e.g.,
$$
(\omega_{kL}-\omega_{k'L})g_{L:L}^{(2)}(k,k')=u_{k'L}\left( z_\sigma 
g_L^{(1)}(k)
+We^{-i\varphi}\int dk'' g_{L:R}^{(2)}(k,k'')u_{k'' R}\right)
$$
and the boundary 
condition: $a_{k\sigma r}(t) \exp(i\omega _{kr}t/\hbar ) \rightarrow 
\beta_{k\sigma r}$ (as $t\rightarrow -\infty $) gives
$$
g_{L:L}^{(2)}(k,k')={u_{k'L}\over \omega_{kL}-\omega_{k'L}-i0}\left( 
z_\sigma g_L^{(1)}(k)
+We^{-i\varphi}\int dk'' g_{L:R}^{(2)}(k,k'')u_{k'' R}\right) \ .
$$
In the same way, a set of linear equations among six functions
$g_{r}^{(1)}(k)$, $g_{r: r'}^{(2)}(k,k')$ ($r,r'=L,R$) is derived and 
its solution 
leads to 
\begin{align}
\beta _{k\sigma r}=a_{k\sigma r}
&+\frac{\chi _{\sigma}(\omega _{kr})}{\Lambda_{\sigma}(\omega _{kr})}
u_{kr}z_{\sigma }f_{\sigma}  \notag \\ 
&+\int dk' \left( \frac{z_{\sigma }^{2}u_{kr}u_{k'r} }
{\omega _{kr}-\omega _{k'r}-i0} 
\frac{\xi_{\sigma}(\omega _{kr})}{\Lambda_{\sigma}(\omega _{kr})} 
a_{k'\sigma r} 
+\frac{z_{\sigma } ^{2}u_{kr}u_{k'\bar r} }
{\omega _{kr}-\omega _{k'\bar r}-i0}
\frac{\kappa _{\sigma}(\omega _{kr})}{\Lambda_{\sigma}(\omega _{kr})} 
a_{k'\sigma \bar r}
\right),
\label{field}
\end{align}
where the notations are the same as those of (\ref{original1}) and 
(\ref{original2}).
On the other hand, because of the completeness of $\beta_{k\sigma r}$ 
which is guaranteed
by the absence of the bound states for $H_F$, one has
$$
f_\sigma =\sum_{r} \int dk [f_\sigma, \beta^\dag_{k\sigma r}]_+ \ 
\beta_{k\sigma r} \ ,
$$
which gives (\ref{original1}). Eq.(\ref{original2}) can be derived 
similarly.}

\begin{thebibliography}{99}
\bibitem{fnote1} Strictly speaking, since the system is infinitely
extended, local equilibrium states of the whole system can not described 
by the density matrices \cite{Bratteli}. However, for the sake of simpler 
presentations, we use such notations.
\bibitem{fnote2} In contrast to the usual scattering problem, the 
perturbed Hamiltonian $H_F$ does not have bound states, but the unpertubed one $H_0$ does.
\bibitem{fnote3} At first glance, the result of ref.\citen{Dong2} does not agree with the present one.
However, this discrepancy is an artifact because of the difference of the unit of the bias voltage.
The authors of ref.\citen{Dong2} measure the voltage in the unit of the Kondo temperature which
increases faster than the peak width as $n$ decreases. In the same unit, the two results agree
well.
\end{thebibliography}
\end{document}